\documentclass{aa}  

\usepackage{graphicx}
\usepackage{txfonts}
\usepackage[colorlinks=true,allcolors=blue]{hyperref}
\usepackage{soul}
\begin{document}

   \title{High N/O ratio at high redshift as a result of a strong burst of star formation and differential galactic winds}

   \author{F. Rizzuti\inst{1,2,3}
          \and
          F. Matteucci\inst{1,2,3}
          \and
          P. Molaro\inst{2,4}
          \and
          G. Cescutti\inst{1,2,3}
          \and
          R. Maiolino\inst{5,6,7}
          }

   \institute{Dipartimento di Fisica, Università degli Studi di Trieste, via Tiepolo 11, I-34143 Trieste, Italy\\
              \email{federico.rizzuti@inaf.it}
         \and
             INAF, Osservatorio Astronomico di Trieste, via Tiepolo 11, I-34143 Trieste, Italy
        \and
        INFN, Sezione di Trieste, via Valerio 2, I-34134 Trieste, Italy
        \and
        Institute for Fundamental Physics of the Universe, via Beirut, 2, I-34151 Trieste, Italy
        \and
         Kavli Institute for Cosmology, University of Cambridge, Madingley Road, Cambridge CB3 0HA, UK
         \and
         Cavendish Laboratory, University of Cambridge, 19 JJ Thomson Avenue, Cambridge CB3 0HE, UK
         \and
         Department of Physics and Astronomy, University College London, Gower Street, London WC1E 6BT, UK
             }

   \date{Received ; accepted }

 
  \abstract
   {Recent observations by JWST have revealed supersolar $^{14}$N abundances in galaxies at very high redshift. On the other hand, these galaxies show subsolar metallicity. The observed N/O ratios are difficult to reproduce in the framework of chemical evolution models for the Milky Way.}
   {Our aim is to reproduce these high N/O ratios with chemical evolution models, assuming different histories of star formation triggering galactic winds, coupled with detailed nucleosynthesis prescriptions for $^{14}$N, $^{12}$C, $^{16}$O, and $^{56}$Fe.}
   {We computed several models for small galaxies ($10^{9}\text{ - }10^{10}\text{ M}_{\odot}$) with a high star formation efficiency and strong galactic winds. These winds are assumed to be differential, mainly carrying out the products of the explosion of core-collapse supernovae. }
   {We find that only models with high star formation rates, a normal initial mass function, and differential galactic winds can reproduce the observed chemical abundances. We also find that, with the same assumptions about star formation and galactic winds, but with a very rapid formation resulting from fast gas infall, we can also reproduce the estimated ages of these objects. We find no necessity to invoke peculiar nucleosynthesis from population III stars, very massive stars, and supermassive stars.}
   {}


   \maketitle
%

\section{Introduction}
Recent data from JWST have indicated supersolar N/O ratios in high-redshift galaxies \citep{2023A&A...677A.115C, 2023ApJ...959..100I, 2024arXiv240702575C, 2024MNRAS.535..881J, 2024A&A...681A..30M, 2024A&A...687L..11S, 2024ApJ...966...92S}.
These high ratios are difficult to reconcile with standard stellar nucleosynthesis and chemical evolution models. 
N is basically a secondary element produced during the CNO cycle, mainly in low- and intermediate-mass stars (LIMSs; 0.8 $\le\text{ M/M}_{\odot}\le$ 8), but it can also be produced as a primary element during the thermal pulses combined with the third dredge-up in asymptotic giant branch (AGB) stars (\citealp{1981A&A....94..175R, 2013MNRAS.431.3642V}). Massive stars also produce some secondary $^{14}$N, which can also be primary if the massive stars rotate \citep{2002A&A...390..561M, 2007A&A...461..571H, 2012A&A...537A.146E} due to the rotation-induced mixing between the He-burning core and H-burning shell that triggers the production of primary $^{14}$N. The fact that $^{14}$N should be produced in a primary fashion by massive stars was first suggested by \cite{1986MNRAS.221..911M}, in order to reproduce the observed N/Fe ratios in halo stars \citep[see also][]{2005A&A...437..429C} and later to explain the N/O abundances in damped Lyman-alpha (DLA) Systems \citep[e.g.][]{1997A&A...321...45M}. In fact, DLA galaxies show the best evidence of the primary nature of N at low metallicities  by displaying a constant N/O ratio with increasing oxygen abundance \citep{1995ApJ...451..100P, 1996A&A...308....1M, 2004oee..sympE..39M, 2003ASPC..304..221M, 2006cams.book..256M, 2003A&A...403...55C, 2014MNRAS.444..744Z, 2018MNRAS.477...56V}. However, even assuming an early primary production of N from massive stars and with the star formation rate (SFR) typical of the Galaxy, it is not possible to reach supersolar N/O ratios at early times. Therefore, the high-$z$ galaxies observed by JWST should have experienced a quite different chemical history relative to the Galaxy. \cite{1997A&A...321...45M} showed that galaxies of different morphological types should present different N/O ratios according to their star formation history. In particular, they showed that high N/O ratios can be obtained in the framework of a chemical evolution model assuming short but intense bursts of star formation, together with a primary production of N by massive stars and differential galactic winds. Differential winds are outflows preferentially carrying out some elements, such as the elements produced by core-collapse supernovae (CC-SNe). 
On the other hand, elements such as N, which are produced mainly by LIMSs, or Fe, mainly originating in type Ia SNe, formed in isolation, have a lower probability of being ejected from the galaxy relative to elements, such as O and $\alpha$-elements, that are mainly produced in massive stars, which tend to cluster together and explode as CC-SNe (see \citealp{2025arXiv250116866B} for a review on star formation). Differential galactic winds have been invoked to reproduce the properties of dwarf galaxies \citep[e.g.][]{1993A&A...277...42P, 1994MNRAS.270...35M, 1995A&A...297..634K} and have been theoretically confirmed by chemo-dynamical simulations \citep{1999ApJ...513..142M, 1999MNRAS.309..941D, 2001MNRAS.322..800R, 2008A&A...489..555R, 2004ApJ...613..159F}. It is reasonable to assume that the gas ejected from a galaxy is the one from CC-SNe, which tend to cluster together. Moreover, differential galactic winds have been also observed by \cite{2002ApJ...574..663M} and \cite{2005MNRAS.358.1453O}. The differential winds should in principle act in any galaxy affected by winds. \cite{2016MNRAS.458.3466V} concluded that differential winds seem to also be necessary in local star-forming galaxies to reproduce the observed high N/O ratios at high O/H. Additionally, the effect of fast-rotating massive stars, and in particular of their winds, on the N/O ratio in the Milky Way stars has been shown by \cite{2010A&A...515A.102C}, although they could not reproduce the very high N/O observed in some halo stars.\\ 
Explaining the supersolar N/O measured by JWST in high-$z$ galaxies is challenging. Recently, \cite{2024ApJ...962L...6K} suggested that the supersolar N abundance observed by JWST in GN-z11 ($z=10.6$) results from intermittent bursts of star formation separated by quiescent periods, thanks to the nucleosynthesis by Wolf-Rayet (WR) stars immediately after the second burst. They also showed that changing the initial mass function (IMF) or including very massive stars and pair-instability SNe is not enough to reproduce the high N/O ratio in single burst models. \\
Also other studies in the literature rely on WR stars as fast producers of high N/O \citep[e.g.][]{2024ApJ...962...50W, 2024A&A...690A.269R}. However, they are confronted with the problem that WR stars produce low O/H, while CC-SNe produce instead high O/H but low N/O. On the other hand,  \cite{2024MNRAS.534.2086T} excluded  WR stars on the basis of the high C/O ratio produced. \cite{2024A&A...688A.142N} have only tested WR stars with solar metallicity, and their preferred sources of high N/O are fast rotators. \cite{2024arXiv241206517S} have recently measured N/O in high-$z$ objects, finding both overabundant and normal values, and also suggest WR stars as possible producers, or alternatively dilution by infall of pristine gas or type II SN winds clearing oxygen before N production.\\
Other studies suggest alternative scenarios to explain the high N/O. \cite{2023A&A...680L..19D} reproduced the observations of GN-z11 with AGB stars \citep{2013MNRAS.431.3642V}, assuming a central accreting black hole and infall of pristine gas. \cite{2024A&A...691A.284R} have shown that the abundances of GN-z11 can be reproduced by assuming a pre-enrichment by massive population III stars in individual clumps with parametrized dilution factors.\\
Other suggestions relative to GN-z11 claimed very massive (>$100\text{ M}_\odot$) and supermassive (>$1000\text{ M}_\odot$) stars as N producers \citep{2023A&A...673L...7C, 2023ApJ...949L..16N, 2024ApJ...966...92S}. Additionally, \cite{2024Natur.627...59M} point out that the nitrogen enrichment of GN-z11 is likely restricted to the very nuclear region of the galaxy, where gas densities exceed $10^8$ cm$^{-3}$ (likely in the vicinity of the black hole).\\
This paper is organized as follows. In Section 2, we describe the JWST observations. In Section 3, we discuss the adopted chemical evolution models and their main parameters. In Section 4, we discuss the assumed nucleosynthesis prescriptions. In Section 5, the results of models assuming a different infall mass, SFR, and galactic wind rates are presented and compared to the observations. Finally, in Section 6 some conclusions are drawn.

\section{Observations}\label{sec:2}
\cite{2023A&A...677A.115C} have first presented the N/O and C/O abundances constrained for the high-redshift galaxy GN-z11 at metallicity $\log(\text{O/H})+12=7.82$ with the unexpectedly high limits of $\log(\text{N/O})>-0.25$ and $\log(\text{C/O})>-0.78$ compared to low-redshift galaxies. Since then, additional measurements for GN-z11 and other high-$z$ galaxies from JWST have been published (see below).\\
In Table \ref{tab:gal}, the complete list of JWST measurements that we have used in this paper to constrain the evolutionary history of high-$z$ galaxies are presented, taken from \cite{2024A&A...681A..30M}, \cite{2023ApJ...959..100I}, \cite{2024ApJ...966...92S}, and \cite{2024arXiv240702575C}. These data have been integrated with previous measurements by \cite{2004MNRAS.355.1132V}, \cite{2012MNRAS.427.1973C}, and \cite{2023ApJ...957...77P}. The galaxies have all a subsolar metallicity of $7<\log(\text{O/H})+12<8.5$, solar or subsolar values of C/O, but supersolar N/O (except GS-z9-0). The Sun is included for comparison. \\
A different case is the one of galaxy GS\_3073, which has recently been measured to have the highest N/O ever found by JWST \citep{2024MNRAS.535..881J}. We include this object in Table \ref{tab:gal}: its $\log(\text{N/O})$ is more than 0.5 dex larger than what was found in the other galaxies. We believe that in order to reach such high N/O, this galaxy must be more evolved than the others. Unfortunately, the age and star formation of GS\_3073 are not constrained \citep{2024MNRAS.535..881J}. For these reasons, we treat this object separately.\\
Some studies also provide estimates for the age and other properties of the high-$z$ galaxies. We list in Table \ref{tab:age} the ages and SFRs available for high-$z$ galaxies, taken from \cite{2024A&A...681A..30M}, \cite{2023A&A...677A..88B}, and \cite{2024arXiv240702575C}. These estimates have been obtained through spectral energy distribution (SED) fitting, based on stellar population synthesis. It is true that these estimates are dependent on the specific choices selected during the fitting, such as the assumed IMF; however, the large error bars of the estimates in Table \ref{tab:age} are often representative of a fitting run with a different IMF, in other words\ \cite{1955ApJ...121..161S} and \cite{2003PASP..115..763C} for \cite{2023A&A...677A..88B}, \cite{2001MNRAS.322..231K} and \cite{2003PASP..115..763C} for \cite{2024arXiv240702575C}, and \cite{2003PASP..115..763C} for \cite{2024A&A...681A..30M}.\\

\begin{table*}
\centering
\footnotesize
\caption{Observational data of JWST high-redshift galaxies (plus the Sun): galaxy name, redshift, metallicity, N/O, C/O, and source.}\label{tab:gal}
\begin{tabular}{cccccc}
\hline\\[-0.7em]
Galaxy&$z$&$\log{(\text{O/H})}+12$&$\log{(\text{N/O})}$&$\log{(\text{C/O})}$&source\\
\\[-0.7em]\hline\\[-0.7em]
GN-z11&10.6&$7.84^{+0.06}_{-0.05}$&$-0.38^{+0.05}_{-0.04}$&---&1, 2\\[0.2em]
GS-z9-0&9.43&$7.49\pm0.11$&$-0.93\pm0.24$&$-0.90\pm0.12$&3\\[0.2em]
CEERS-1019&8.68&$7.70\pm 0.18$&$-0.18\pm0.11$&$-0.75\pm0.11$&2\\[0.2em]
ERO\_04590&8.50&$7.19^{+0.13}_{-0.10}$&---&$-0.53^{+0.17}_{-0.15}$&4\\[0.2em]
GLASS\_150008&6.23&$7.65^{+0.14}_{-0.08}$&$-0.40^{+0.05}_{-0.07}$&$-1.08^{+0.06}_{-0.14}$&4\\[0.2em]
CEERS\_00397&6.00&$7.99^{+0.12}_{-0.10}$&---&$-0.65^{+0.14}_{-0.12}$&4\\[0.2em]
GS\_3073&5.55&$8.00^{+0.12}_{-0.09}$&$0.42^{+0.13}_{-0.10}$&$-0.38^{+0.13}_{-0.11}$&5\\[0.2em]
GLASS\_150029&4.58&$7.73^{+0.09}_{-0.08}$&---&$-0.84^{+0.10}_{-0.09}$&4\\[0.2em]
SMACS2031&3.51&$7.76 \pm 0.1$&$-0.66 \pm 0.1$&$-0.80 \pm 0.09$&2, 6\\[0.2em]
Lynx arc&3.36&$7.87 \pm 0.2$&$-0.53 \pm 0.2$&$-0.14 \pm 0.2$&2, 7\\[0.2em]
Sunburst&2.37&$8.03\pm 0.06$&$-0.21^{+0.10}_{-0.11}$&$-0.51\pm0.05$&2, 8\\[0.2em]
Sun&0&8.69&$-0.86$&$-0.26$&9\\
\\[-0.7em]\hline\\[-0.7em]
\multicolumn{6}{l}{\begin{minipage}{0.57\textwidth}Sources. 1: \cite{2024ApJ...966...92S}, 2: \cite{2024A&A...681A..30M}, 3: \cite{2024arXiv240702575C}, 4: \cite{2023ApJ...959..100I}, 5: \cite{2024MNRAS.535..881J}, 6: \cite{2012MNRAS.427.1973C}, 7: \cite{2004MNRAS.355.1132V}, 8: \cite{2023ApJ...957...77P}, 9: \cite{2009ARA&A..47..481A}.\end{minipage}}
\end{tabular}
\end{table*}
\begin{table}
\centering
\footnotesize
\caption{Observational data of JWST high-redshift galaxies: galaxy name, estimated age, SFR, and source.}\label{tab:age}
\begin{tabular}{cccc}
\hline\\[-0.7em]
Galaxy&age (Myr)&SFR (M$_\odot$ yr$^{-1}$)&source\\
\\[-0.7em]\hline\\[-0.7em]
CEERS-1019 A&$4.0\pm0.3$&$148\pm25$&1\\[0.2em]
CEERS-1019 B&$5.7\pm0.7$&$83\pm18$&1\\[0.2em]
GN-z11&$18.6^{+10.2}_{-5.4}$&$25\pm5$&2\\[0.2em]
GS-z9-0&$32^{+20}_{-9}$&$5.46\pm1.04$&3\\
\\[-0.7em]\hline\\[-0.7em]
\multicolumn{4}{l}{\begin{minipage}{0.8\columnwidth}Sources. 1: \cite{2024A&A...681A..30M}, 2: \cite{2023A&A...677A..88B}, 3: \cite{2024arXiv240702575C}.\end{minipage}}
\end{tabular}
\end{table}

\section{The chemical evolution model}
The adopted chemical evolution model assumes one intense and short starburst occurring at the very beginning of the galaxy evolution. This model is similar to the one described in \cite{1997A&A...321...45M}, but is much more detailed for what concerns stellar nucleosynthesis.
The model assumes that the studied galaxy forms by infall of primordial gas, with a strong SFR triggering galactic winds, which can be normal or differential. Core-collapse supernovae (CC-SNe) explode before SNe Ia and are generally clustered together, triggering a differential wind with different chemical elements lost from the galaxy at different rates \citep{1993A&A...277...42P, 1994MNRAS.270...35M, 1995A&A...297..634K}. Therefore, it is plausible that the $\alpha$-elements, such as oxygen, are lost more easily from the galaxy than other elements such as nitrogen that are mainly produced in LIMSs.
Differential winds have been studied theoretically by several authors \citep{1999ApJ...513..142M, 1999MNRAS.309..941D, 2001MNRAS.322..800R, 2004ApJ...613..159F} and also observed \citep[see][and references therein]{2008A&A...489..555R}.\\
The SFR depends on the gas mass:
\begin{equation}
\psi(t) = -\nu\, M_\text{gas}(t)
,\end{equation}
where $\nu$ is the efficiency of star formation and is expressed in units of Gyr$^{-1}$. The assumed IMF is the one of \cite{1955ApJ...121..161S}. \\
The infall rate, $I_{i,\text{inf }}(t)$, corresponding to element $i$ is an exponential one:
\begin{equation}
 I_{i,\text{inf }}(t)= a\,X_{i,\text{inf }}\, e^{-t/\tau}   
,\end{equation}
where $X_{i,\text{inf }}$ is the abundance of the element $i$ in the infalling gas, which is assumed to be primordial and therefore containing no metals. The quantity $\tau$ is the timescale of the mass accretion. The quantity $a$ is a constant that is derived by assuming that a given infall mass is reached at the present time (13.7 Gyr).\\
The rate of galactic wind, $W_i(t)$, corresponding to element $i$ is proportional to the SFR:
\begin{equation}
    W_i(t) = -\omega_i\, \psi(t)\, X_{i,\text{w }}(t)
,\end{equation}
where $\omega_i$ is the mass loading factor and is the same for all the chemical elements for `classical' wind, whereas it varies from element to element if a differential wind is assumed. The abundance, $X_{i,\text{w}}$, is relative to the element $i$ in the wind. In particular, the mass loading factor is equal to 1 for O and the other $\alpha$-elements, almost entirely produced by CC-SNe, to 0.7 for C, produced mostly by massive stars, to 0.3 for Fe, produced one third by CC-SNe and two thirds by SNe Ia, and to 0 for the other elements. \\
The model relaxes the instantaneous recycling approximation and the equations for each element are solved numerically.
We include a detailed treatment of the SN rates. In particular, for the type Ia SN rate, we adopt a delay distribution function related to the single degenerate scenario that, together with the double degenerate scenario, better reproduces the observations of chemical abundances (see \citealp{2021A&ARv..29....5M} for a review) as well as the cosmic type Ia SN rate \citep[see][]{2024A&A...689A.203P}. The model also includes the calculation of the nova and merging neutron star rates, described in \cite{2023MNRAS.523.2974M}. However, novae and merging neutron stars produce elements that are not treated in this paper, in which we concentrate on $^{14}$N and $^{16}$O.\\
Before discussing in detail the assumed prescriptions for stellar nucleosynthesis, we remind the reader here that $^{14}$N is mainly produced in LIMSs but that a fraction of it arises from massive stars, while $^{16}$O is entirely produced by massive stars. Moreover, while $^{16}$O is a primary element, $^{14}$N is secondary but a part of it can be produced as a primary both in LIMSs and massive stars. Finally, $^{12}$C is a primary element mainly produced by rotating massive stars and in part by LIMSs (see next section).

\begin{table*}
\centering
\footnotesize
\caption{Model parameters: model number, galaxy mass, massive star yields (WW95: \citealp{1995ApJS..101..181W}; LC: \citealp{2018ApJS..237...13L}), wind parameter, $\omega$, differential wind, star formation efficiency, $\nu$, infall timescale, $\tau$, and star formation burst duration, $\Delta t_{SF}$.\\}\label{tab:mod}
\begin{tabular}{cccccccc}
\hline\\[-0.7em]
model&mass&massive star&$\omega$&differential&$\nu$&$\tau$&$\Delta t_{SF}$\\
&(M$_\odot$)&yields&&wind&(Gyr$^{-1}$)&(Gyr)&(Myr)\\
\\[-0.7em]\hline\\[-0.7em]
0&$10^9$&WW95&80&yes&30&0.5&150\\
1&$10^9$&LC 150 - 000&80&yes&1.5&0.5&200\\
2&$10^9$&LC 150 - 000&80&yes&30&0.5&150\\
3&$10^9$&LC 150 - 000&0&no&30&0.5&200\\
4&$10^9$&LC 150 - 000&700&yes&300&0.5&100\\
5&$10^{10}$&LC 150 - 000&80&yes&30&0.5&150\\
6&$10^9$&LC 150 - 000&80&no&30&0.5&250\\
7&$10^9$&LC 150 - 000&80&yes&15&0.01&100\\
8&$10^9$&LC 150 - 000&1500&yes&80&0.0001&25\\
\\[-0.7em]\hline
\end{tabular}
\end{table*}

\section{Nucleosynthesis prescriptions}
It is now established that nitrogen can be produced as a primary element in metal-poor massive stars, as was first suggested by \cite{1986MNRAS.221..911M} and later by \cite{1997A&A...321...45M}, once stellar rotation is taken into account, as was confirmed by the chemical evolution models of \cite{2006A&A...449L..27C, 2008A&A...479L...9C}, which employed the rotating models for massive stars by \cite{ 2002A&A...390..561M, 2002A&A...381L..25M} and \cite{2007A&A...461..571H}. This happens because rotation enhances the mixing between layers, so during He-burning carbon is transported outside the convective core into the H-burning shell, where it is converted into nitrogen and later brought to the surface. However, for a substantial production of nitrogen, stars are required to have very fast rotation and low metallicity, which enhances the effects of rotation. \\
Since these pioneering studies, much progress has been made in modelling the evolution of rotating stars \citep[e.g.][]{2013ApJ...764...21C, 2016MNRAS.456.1803F, 2024ApJS..270...28R}. Recently, the work of \cite{2024A&A...688A.142N} has also shown the importance of low-metallicity fast rotators to explain the N production in high-$z$ galaxies. In particular, the work of \cite{2018ApJS..237...13L} provides a grid of stellar models with different mass, metallicity, and rotation. This grid represents an excellent basis for chemical evolution models of galaxies, which need to include a large range of stars with different physical properties as producers. Specifically, the possibility of including stars that rotate at different velocities proved to be very interesting for investigating the effects of stellar rotation on the nucleosynthesis. For these reasons, the stellar models of \cite{2018ApJS..237...13L} have been systematically used for Galactic archaeology by independent research groups over the years \citep{2018MNRAS.476.3432P, 2019MNRAS.490.2838R, 2019MNRAS.489.5244R, 2019MNRAS.489.3539G, 2022IAUS..366...63K, 2023MNRAS.523.2974M}. \\
In this paper, we have used the grid of models from \cite{2018ApJS..237...13L} for the nucleosynthesis from massive stars. It is worth summarizing here their main features. The grid of stellar models consists of nine masses between 13-120 M$_\odot$, four metallicities, [Fe/H] = 0, --1, --2, and --3, and three different rotations with an initial equatorial velocity of 0 (non-rotating), 150, and 300 km s$^{-1}$. \cite{2018ApJS..237...13L} also present different sets of yields based on different physical assumptions. Here, we use Set R (recommended), which assumes a mass cut for envelope ejection chosen to produce 0.07 M$_\odot$ of $^{56}$Ni, and an explosion only for stars with masses $\leq25$ M$_\odot$, while masses $>25$ M$_\odot$ only contribute to the yields via stellar winds. This assumption is supported by both observational \citep{2009MNRAS.395.1409S} and theoretical studies \citep{2016ApJ...821...38S, 2016MNRAS.460..742M} of CC-SNe. \\
Since we can choose from three different rotations in the grid of massive stars, we need to select a velocity distribution. In recent years, many studies have been investigating the question of stellar rotation in the context of galactic archaeology \citep{2018MNRAS.476.3432P, 2019MNRAS.489.5244R, 2021MNRAS.502.2495R, 2019MNRAS.490.2838R, 2023MNRAS.523.2974M}. They all agree that, in order to explain the chemical history of the Milky Way, massive stars are expected to rotate faster at lower metallicity, but have very low rotation at solar metallicity. Inspired by these works, we decided to give massive stars an initial equatorial rotation of 150 km s$^{-1}$ for [Fe/H] $\leq$ --2 dex, and 0 km s$^{-1}$ for [Fe/H] > --2 dex. \cite{2018MNRAS.476.3432P} show that a smoother transition is more physical, and \cite{2021MNRAS.502.2495R} that a stochastic distribution of velocities is preferable; however, a simplified distribution such as the one we assume here is accurate enough to reproduce the evolution of light elements, as is shown for example in \cite{2019MNRAS.490.2838R}. Choosing a different rotation velocity distribution would change the initial N/O plateau, which is still unconstrained in high-$z$ galaxies, but would not have an effect at higher metallicity in our models, which are mostly driven by differential winds, since just changing the rotation in massive stars cannot explain the supersolar N/O.\\
With these assumptions, massive stars above $\sim40$ M$_\odot$ likely enter the WR stage, producing an enrichment only by stellar wind often comparable to the one from CC-SNe. In particular, \cite{2018ApJS..237...13L} show that N can be produced both in WR and CC-SNe, but O is always produced orders of magnitude more in CC-SNe. Furthermore, both WR and CC-SNe are expected to form in clustered regions, so they would both be subject to ejection by differential winds.\\
In this present paper, in order to show the difference from previous models with non-rotating massive stars, we present results obtained with the non-rotating grid of massive star models by \cite{1995ApJS..101..181W}.\\
Finally, the nucleosynthesis at higher metallicity also needs the contribution from LIMSs, which in the Milky Way becomes dominant around [Fe/H] > --2 dex. The nucleosynthesis for LIMSs (< 8 M$_\odot$) in our chemical evolution model is assumed from the grid of stellar models by \cite{2010MNRAS.403.1413K}. However, our models for high-redshift galaxies presented here only show the evolution up to < 200 Myr, which means that stars below 4-5 M$_\odot$ do not have time to enrich the interstellar medium. Their contribution is therefore negligible, relative to the results shown.

\section{Results}
We ran several models by changing the efficiency of star formation, the burst duration, the galactic wind efficiency (mass loading factor), and the infall timescale. All the models have the same IMF, which is the standard \cite{1955ApJ...121..161S} one\footnote{$x=1.35$, in the range 0.1 - 100 M$_{\odot}$}, and differential galactic winds, where only the products of CC-SNe are expelled from the galaxy, except in Models 3 and 6, which assume a classical galactic wind whereby all the elements are expelled at the same rate.
In Table \ref{tab:mod}, we show the parameters of the models. In particular, in column 2 we show the assumed infall mass, in column 3 we report the adopted prescriptions for the stellar yields, in column 4  there is the mass loading factor, applied only to selected elements, as was described before, in column 5 is indicated the type of wind (classical or differential), in column 6 we report the assumed star formation efficiency in units of Gyr$^{-1}$, in column 7 we find the assumed infall timescale in units of gigayears, and finally in column 8 there is the assumed duration of the burst of star formation. Model 0 is a classical model with old nucleosynthesis prescriptions for massive stars \citep{1995ApJS..101..181W} that can be compared to a similar model in \cite{1997A&A...321...45M}. Models from 0 to 4  and from 6 to 8 refer to an infall mass of $10^{9}\text{ M}_{\odot}$, while Model 5 is relative to a mass of $10^{10}\text{ M}_{\odot}$.
\begin{figure*}
   \centering
   \includegraphics[width=0.49\hsize]{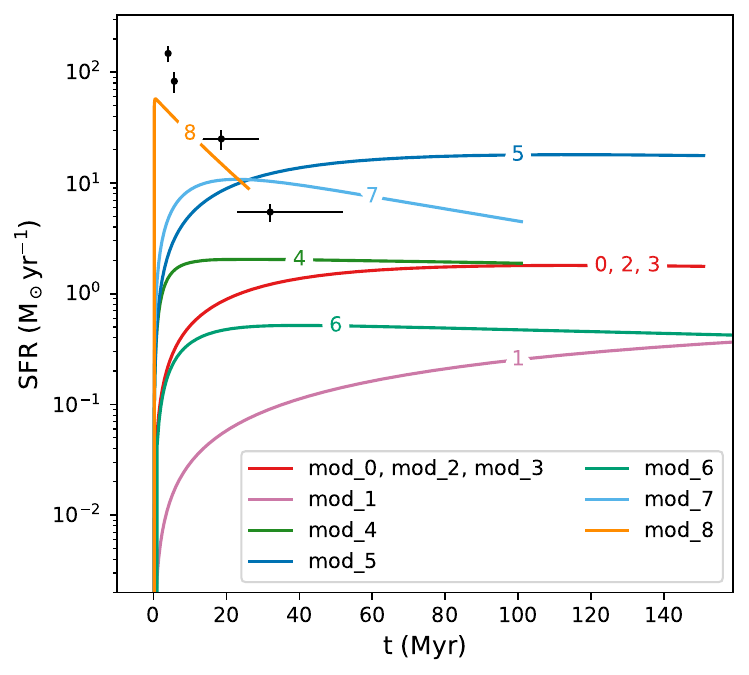}
   \includegraphics[width=0.49\hsize]{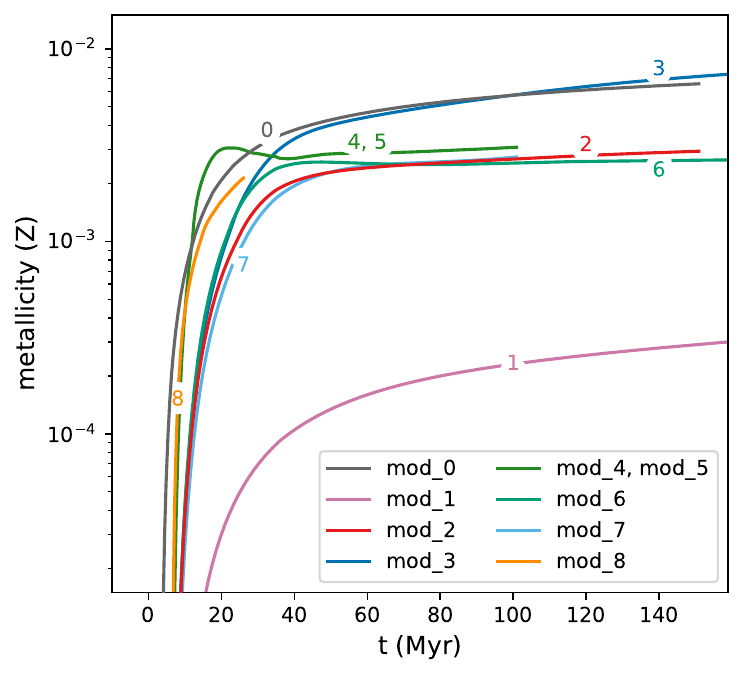}
      \caption{Star formation rate (left) and metallicity (right) as a function of time, as predicted by all models in Table \ref{tab:mod}. The black dots and associated error bars are JWST measurements from \cite{2024A&A...681A..30M}, \cite{2023A&A...677A..88B}, and \cite{2024arXiv240702575C} (see Table \ref{tab:age}).}
         \label{fig:SFR}
   \end{figure*}
   \begin{figure*}
   \centering
   \includegraphics[width=0.49\hsize]{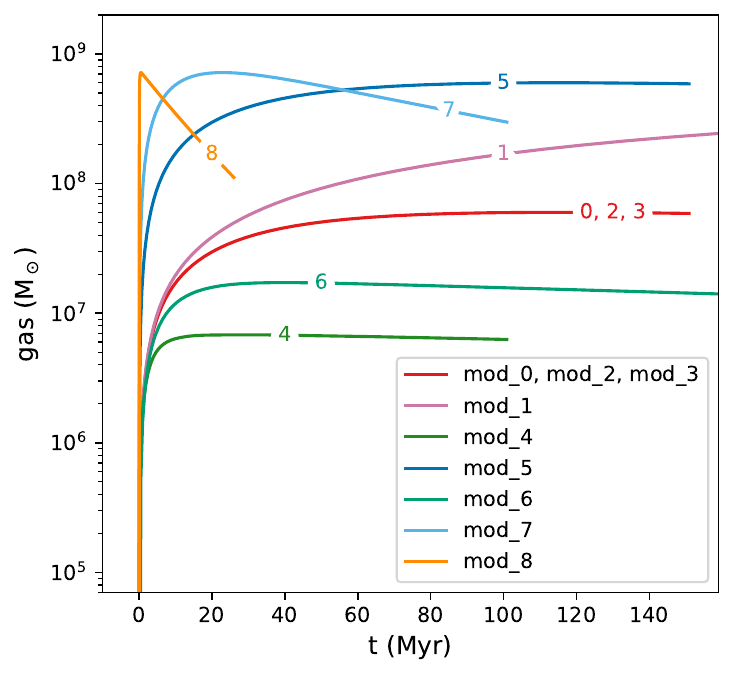}
   \includegraphics[width=0.49\hsize]{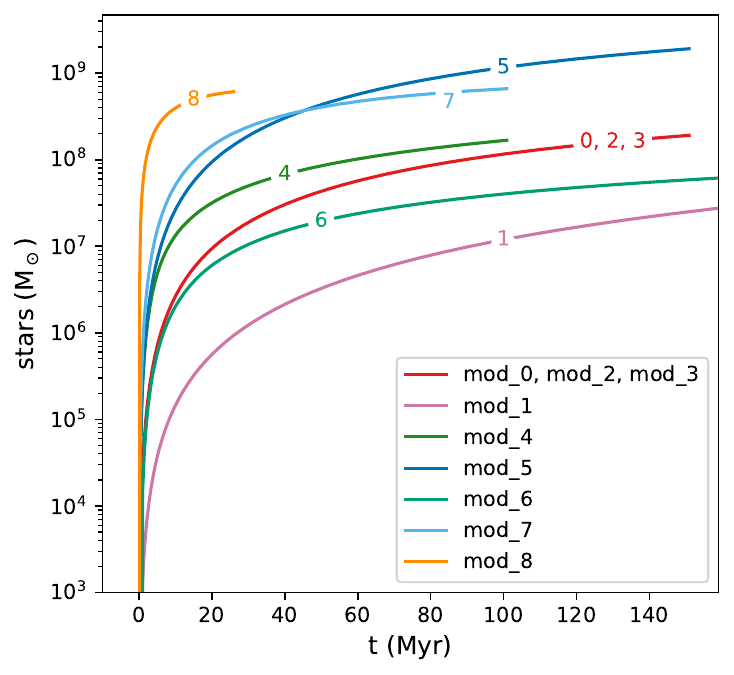}
      \caption{Gas (left) and star (right) mass in the galaxy as a function of time, as predicted by all models in Table \ref{tab:mod}.}
         \label{fig:gas}
   \end{figure*}
\\In Fig.\ \ref{fig:SFR}, we show the SFR versus galactic time for each of the models. Clearly, at the beginning, the SFR of Model 7 and 8 is higher than in all the other cases. The reason for that resides in the fact that they are almost closed-box models, because of the assumed very short infall timescale. In this case, more gas is present at early times relative to the other models with longer infall timescales, and therefore the SFR is higher. Model 5 is the only one referring to a more massive object and its SFR is higher than the rest of the models, which are relative to an object of $10^{9}\text{ M}_{\odot}$ and which all have the same infall timescale. In the same figure (right panel), we show the evolution of the global metallicity, $Z$, for each model. As one can notice, the increase in the gas metallicity  in the first 50 Myr is very fast in all models.
\\In Fig.\ \ref{fig:gas}, we show the predicted evolution of the gas and star mass in each model; while the star mass always increases, the gas mass is often flat or also increasing, in spite of the star formation and galactic wind, and this is because we are looking at the phases of high gas infall (short infall timescale). Only at later times does the gas start to decrease continuously. The only exceptions are Models 7 and 8, in which the infall has even shorter timescales (closed-box models).
   \begin{figure*}
   \centering
   \includegraphics[width=\hsize]{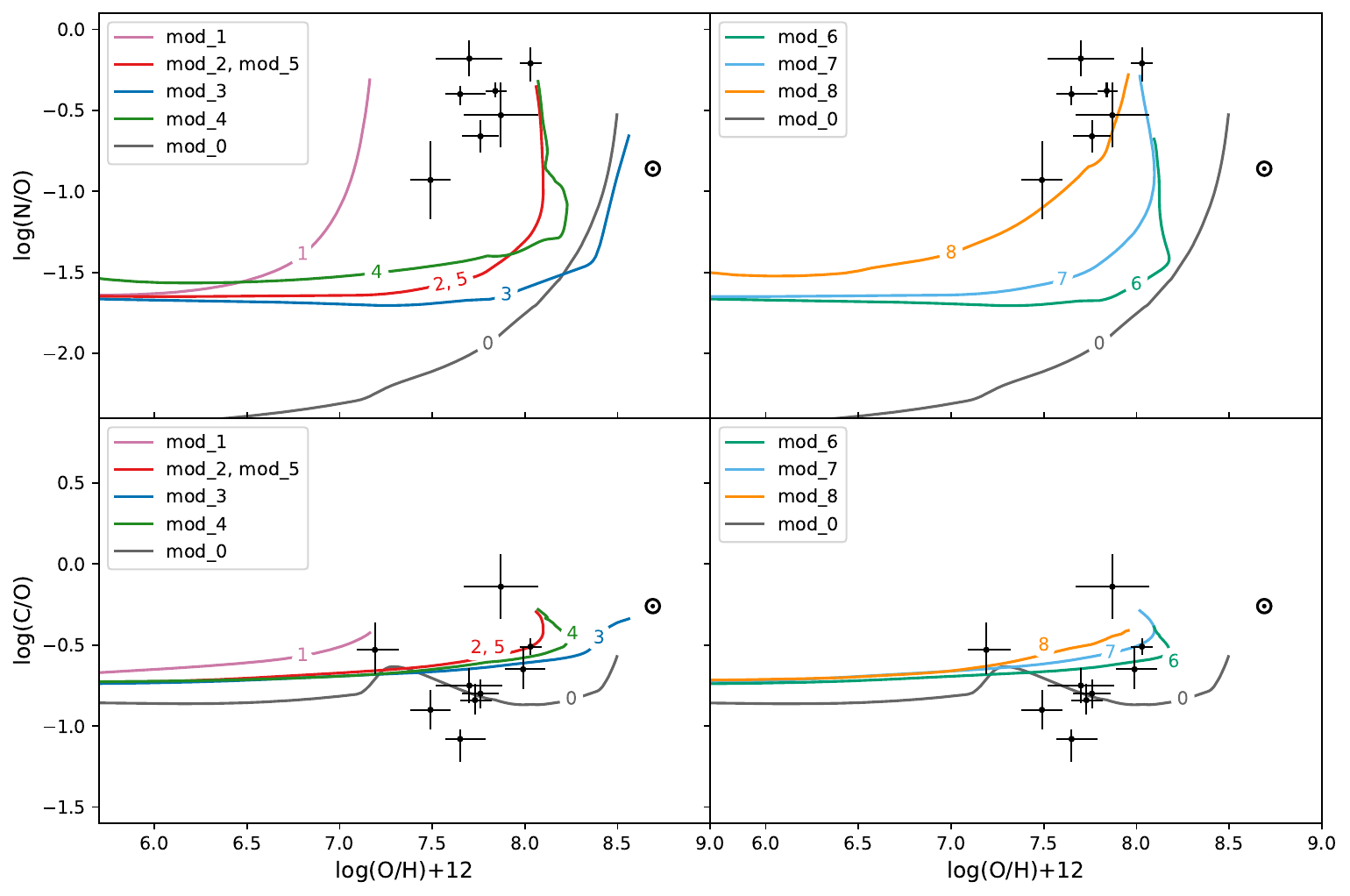}
      \caption{Predicted N/O (top panels) and C/O (bottom panels) versus log(O/H)+12, for all models in Table \ref{tab:mod}. The black dots and associated error bars are JWST measurements from \cite{2024A&A...681A..30M}, \cite{2023ApJ...959..100I, 2024ApJ...966...92S}, and \cite{2024arXiv240702575C} (see Table \ref{tab:gal}).}
         \label{fig:NOCO}
   \end{figure*}
\\In Fig.\ \ref{fig:NOCO}, we report the predicted N/O and C/O ratios from our models compared to the data from JWST by \cite{2024A&A...681A..30M}, \cite{2023ApJ...959..100I}, \cite{2024ApJ...966...92S}, and \cite{2024arXiv240702575C}. We do not include here GS\_3073, which is studied separately. It is clear from the figure that the best models for reproducing the very high N/O ratio are Models 2, 4, 5, 7, and 8. Model 2 and Model 4 have very efficient star formation and galactic winds; Model 7 and Model 8 also have a very high SFR and winds but a very short infall timescale, which simulates a closed-box model, and a shorter duration of star formation. Model 0 predicts too low ratios and shows the necessity of producing primary N at low metallicity. Model 1, although it does not reproduce the data, predicts values of the N/O ratio at low metallicities similar to the other models. It is clear that only models with differential galactic winds can approach the high N/O ratios observed. In fact, Models 3 and 6 with classical winds can only reach solar N/O thanks to the AGB production, but the ratio cannot increase further even when the model is left to evolve for a longer time (200-250 Myr).
\\In Fig.\ \ref{fig:NOCO} are also shown the predicted and observed C/O ratios. Most models can reproduce the C/O ratios, except Model 0, without rotating massive stars, and 1, with lower star formation; also, models without differential winds can reproduce the C/O but cannot reproduce the high N/O. It should be noted that with Models 7 and 8, with a very fast infall rate, the estimated ages of the observed objects are better reproduced (see Table \ref{tab:mod} for comparison). \\
It is clear that to reproduce the high N abundance measured in these high-redshift objects, besides assuming detailed nucleosynthesis prescriptions for N, including its primary and secondary production, it is necessary to assume strong differential galactic winds triggered by a very strong SFR.
These hypotheses are quite reasonable, since a clear correlation has been observed between star formation and galactic winds, especially winds enriched in elements produced by CC-SNe \citep{2002ApJ...574..663M, 2005MNRAS.358.1453O} such as oxygen. We remind the reader that a small fraction of $^{14}$N is also produced and ejected by massive stars and should mainly be produced in a primary fashion \citep[see][]{1986MNRAS.221..911M, 2002A&A...381L..25M}. However, the primary N produced by rotating massive stars is not enough by itself to reproduce the high N abundance observed. Concerning carbon, it is mainly produced by rotating massive stars and partly from LIMSs \citep{2020A&A...639A..37R}. Fe is mainly produced by type Ia SNe (roughly 70 percent for a normal Salpeter-like IMF).
   \begin{figure}
   \centering
   \includegraphics[width=\hsize]{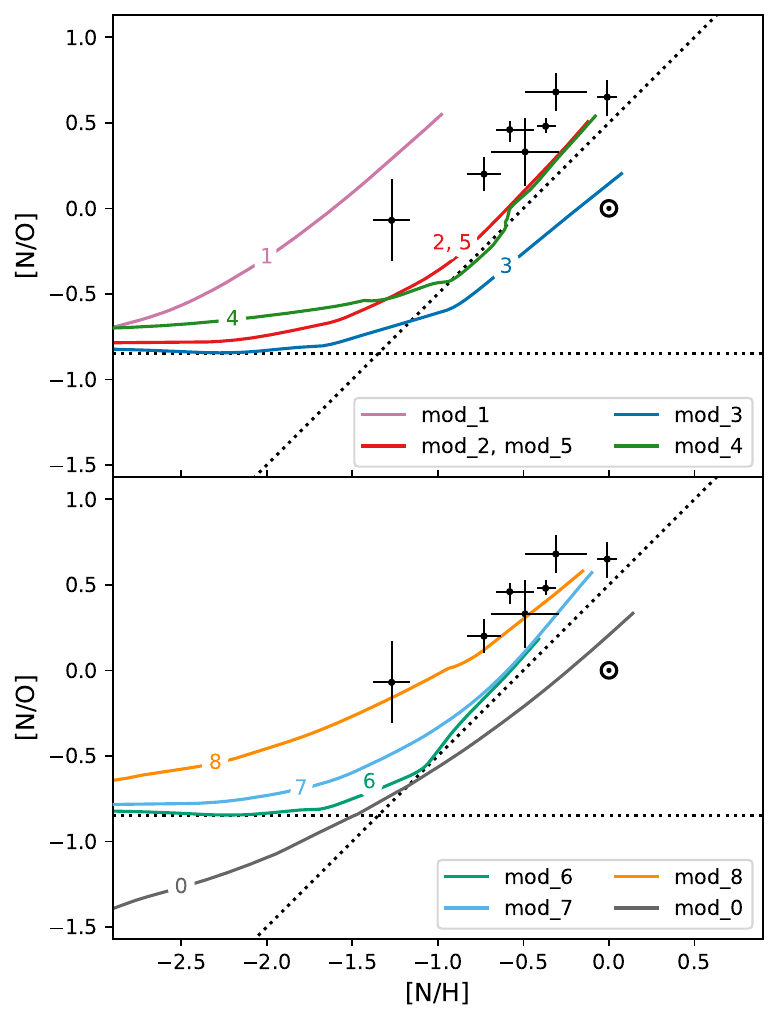}
      \caption{Predicted [N/O] versus [N/H], for all models in Table \ref{tab:mod}. The dotted lines are empirical primary and secondary N production from \cite{2014MNRAS.444..744Z}. The black dots and associated error bars are JWST measurements from \cite{2024A&A...681A..30M}, \cite{2023ApJ...959..100I, 2024ApJ...966...92S}, and \cite{2024arXiv240702575C} (see Table \ref{tab:gal}).}
         \label{fig:NONH}
   \end{figure}
   \begin{figure*}
   \centering
   \includegraphics[width=\hsize]{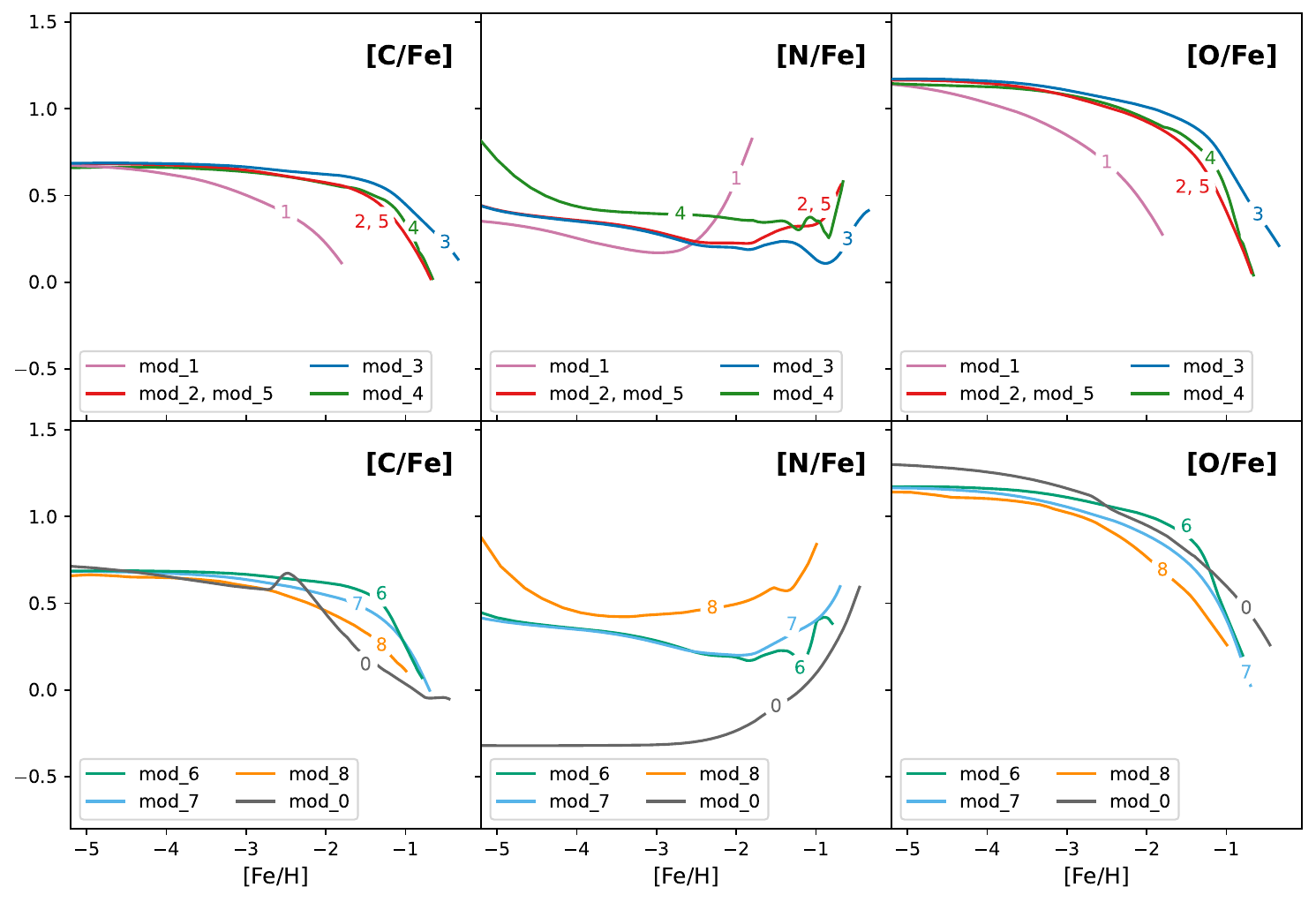}
      \caption{Predicted [C/Fe] (left panels), [N/Fe] (central panels), and [O/Fe] (right panels) versus [Fe/H], for all models (top and bottom panels) in Table \ref{tab:mod}.}
         \label{fig:CNO}
   \end{figure*}
\\ We show in Fig.\ \ref{fig:NONH} the [N/O] versus [N/H] predictions from the models. These ratios show the primary and secondary production of N, as is illustrated by the empirical dotted lines from \cite{2014MNRAS.444..744Z} (see references within). In fact, at low metallicity the models follow a plateau, since N and O are both produced as primary elements from the same sources (massive stars), and therefore they do not depend on metallicity. The only model that does not follow this behaviour is Model 0, since it does not have any primary N production.
\\Finally, in Fig.\ \ref{fig:CNO} we show the [CNO/Fe] ratios predicted by all our models. The [X/Fe] notation is relative to the solar abundances, which are therefore all equal to zero in this notation. In these plots one can notice that at very low metallicities ([Fe/H]$<-3$ dex) the abundance ratios are rather flat. This is because, at these metallicities, only massive stars contribute to both CNO and Fe. It is only after the occurrence of the first type Ia SNe that the ratios start to decrease, because of the injection of Fe. The curves of Fig.\ \ref{fig:CNO} are only predictions at the moment, since no data are yet available.\\
Additionally, these results allow us to also make predictions on the evolution of another important isotope, $^{13}$C. Since the same nucleosynthesis sources that produce $^{14}$N also contribute to $^{13}$C, we expect from the above results that the $^{13}$C/$^{12}$C ratio would fastly increase with metallicity,  reaching supersolar values of $^{13}$C as it also happens to $^{14}$N. Recent detections of $^{13}$C in DLA galaxies at $z\sim 2$ \citep{2017A&A...597A..82N, 2020MNRAS.494.1411W, 2024MNRAS.534...12M} and in Milky Way halo stars \citep[see references within]{2023A&A...679A..72M} also seem to point in this direction. 
\\The very high $\log(\text{N/O})=0.42$ of galaxy GS\_3073, larger than the other high-$z$ galaxies \citep{2024MNRAS.535..881J}, cannot be explained by any of the models presented in Fig.~\ref{fig:NOCO}. We believe that this is because this object must be more evolved compared to the other galaxies. Unfortunately, its age is not estimated. In order to reproduce this galaxy, we extended the evolution of our best models for a longer time. In particular, Models 2, 4, and 5 were run for 1 Gyr, Model 7 for 200 Myr, and Model 8 for 50 Myr. We present the results in Fig.~\ref{fig:gs}: Models 2, 4, and 5 stop around $\log(\text{N/O})\sim0$, while only Models 7 and 8 can reach an N/O compatible with GS\_3073. Models 7 and 8 are also in agreement with the C/O ratio of GS\_3073. We conclude that GS\_3073 can be explained by a very rapid infall of primordial gas, in addition to differential winds, and is in a later evolutionary stage compared to the other high-$z$ galaxies; we expect its age to be larger than $\sim50$ Myr.
   \begin{figure}
   \centering
   \includegraphics[width=\hsize]{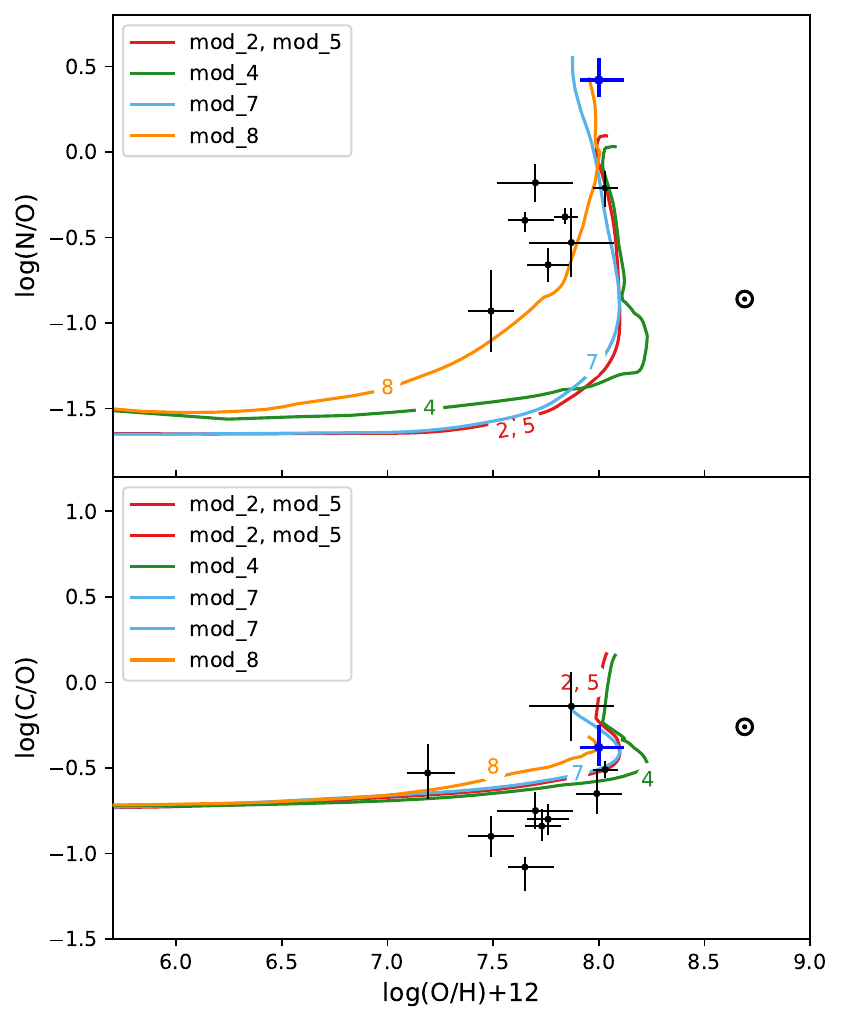}
      \caption{Same as Fig.~\ref{fig:NOCO}, but for JWST measurements of galaxy GS\_3073 (blue dots and error bars) by \cite{2024MNRAS.535..881J}, and Models 2, 4, and 5 evolved for 1 Gyr, Model 7 for 200 Myr, and Model 8 for 50 Myr.}
         \label{fig:gs}
   \end{figure}

\section{Conclusions and discussion}
In this paper, we have computed the chemical evolution of galaxies forming by a fast infall of gas and affected by short and intense starbursts followed by very efficient galactic winds (classical and differential) at a high redshift. In particular, we have followed the evolution of C, N, O, and Fe by means of a chemical evolution model taking into account detailed stellar nucleosynthesis from massive stars (winds and CC-SNe), LIMSs, type Ia SNe, novae, and merging neutron stars. We have assumed a normal Salpeter-like IMF without invoking very massive or supermassive population III stars. 
Our main conclusions can be summarized as follows:
\begin{itemize}
    \item We found that galaxies with an infall mass of $10^{9}\ \text{M}_{\odot}$ and a very intense SFR with an efficiency in the range of 15-300 Gyr$^{-1}$, and differential winds (losing preferentially the products of CC-SNe such as O and $\alpha$-elements) with mass loading factors from 80 to 1500, can reproduce the supersolar N/O ratio observed in high-$z$ galaxies by JWST as well as the ages inferred for these objects from population synthesis models.
    
    \item The model with a higher infall mass of $10^{10}\ \text{M}_{\odot}$, a star formation efficiency of 30 Gyr$^{-1}$, and a differential galactic wind with a mass loading factor of 80 can also reproduce the high observed N/O.
    \item Models with classical galactic wind, even with a very high star formation efficiency and mass loading factor, cannot reproduce the observations.

    \item The models with extremely short infall timescales (10 and 0.1 Myr), besides reproducing the high N abundances, do also agree with the very young estimated ages of the observed galaxies. GS\_3073 can only be explained by these rapid-infall models, assuming it is more evolved than the other high-$z$ galaxies.
   
    \item There is no necessity to invoke peculiar nucleosynthesis from massive population III stars to reproduce the features of the high-$z$ galaxies \citep[see also][]{2024ApJ...962L...6K}.

    \item We predict that $^{13}$C, which is produced by the same stars and in the same fashion as $^{14}$N, would also present supersolar values in high-$z$ galaxies.
    
\end{itemize}
It is interesting to draw a comparison to the other studies in the literature that have tackled the problem of the high N/O ratio in high-$z$ galaxies. As was already discussed, \cite{2024ApJ...962L...6K} are able to explain the JWST data from \cite{2023A&A...677A.115C} with a chemical evolution model assuming an intermittent SF history, specifically two strong starbursts separated by a quiescent phase of 100 Myr. They assume that the main producers of the high N/O are WR stars. Since they do not assume differential galactic winds, they need to postulate intermittent SF in order to reach high N/O and O/H at the same time. In our paper instead, thanks to the differential galactic winds, we are able to reproduce the observations without invoking intermittent star formation.\\
The other studies in the literature that rely on WR stars as fast producers of high N/O \citep[e.g.][]{2024ApJ...962...50W, 2024A&A...690A.269R} are confronted with the problem that WR stars produce low O/H, while instead CC-SNe produce high O/H but low N/O. In this paper, we solve this problem by means of differential galactic winds. Indeed, both WR and AGB stars represent a source of high N/O, but it should be noted that WR stars are rare but are the first to die within few million years, while AGB stars are long-lived but are the dominant enrichment source. \\
As was mentioned before, \cite{2023A&A...680L..19D} reproduced the observations of GN-z11 with AGB stars \citep{2013MNRAS.431.3642V}, assuming a central accreting black hole and infall of pristine gas, and therefore estimated an age of 40 - 130 Myr for this object, which is longer than previous estimates. This explanation is specific to systems with a central accreting black hole.\\
Other works have suggested the contribution from very massive and supermassive stars \citep{2023A&A...673L...7C, 2023ApJ...949L..16N, 2024ApJ...966...92S}, but their properties are still largely unconstrained and our work shows that they are not necessary. We therefore can conclude that, regardless of the nucleosynthesis source, differential galactic winds remain a simple and effective way of explaining the JWST observations.

\begin{acknowledgements}
F.R.\ and G.C.\ acknowledge the grant PRIN project No. 2022X4TM3H “Cosmic POT” from Ministero dell’Università e della Ricerca (MUR). F.M.\ thanks I.N.A.F. for the 1.05.12.06.05 Theory Grant - Galactic archaeology with radioactive and stable nuclei. F.M.\ acknowledges also support from Project PRIN MUR 2022 (code 2022ARWP9C) “Early Formation and Evolution of Bulge and HalO (EFEBHO)” (PI: M. Marconi). R.M. acknowledges support by the Science and Technology Facilities Council (STFC), by the ERC through Advanced Grant 695671 “QUENCH”, and by the UKRI Frontier Research grant RISEandFALL. R.M. also acknowledges funding from a research professorship from the Royal Society. We thank the anonymous referee for the helpful comments that helped us improve this paper.
\end{acknowledgements}

   \bibliographystyle{aa} 
   \bibliography{aa.bib} 

\begin{thebibliography}{81}
\expandafter\ifx\csname natexlab\endcsname\relax\def\natexlab#1{#1}\fi

\bibitem[{{Asplund} {et~al.}(2009){Asplund}, {Grevesse}, {Sauval}, \& {Scott}}]{2009ARA&A..47..481A}
{Asplund}, M., {Grevesse}, N., {Sauval}, A.~J., \& {Scott}, P. 2009, \araa, 47, 481

\bibitem[{{Beuther} {et~al.}(2025){Beuther}, {Kuiper}, \& {Tafalla}}]{2025arXiv250116866B}
{Beuther}, H., {Kuiper}, R., \& {Tafalla}, M. 2025, accepted for Annual Reviews of Astronomy and Astrophysics, arXiv:2501.16866

\bibitem[{{Bunker} {et~al.}(2023){Bunker}, {Saxena}, {Cameron}, {Willott}, {Curtis-Lake}, {Jakobsen}, {Carniani}, {Smit}, {Maiolino}, {Witstok}, {Curti}, {D'Eugenio}, {Jones}, {Ferruit}, {Arribas}, {Charlot}, {Chevallard}, {Giardino}, {de Graaff}, {Looser}, {L{\"u}tzgendorf}, {Maseda}, {Rawle}, {Rix}, {Del Pino}, {Alberts}, {Egami}, {Eisenstein}, {Endsley}, {Hainline}, {Hausen}, {Johnson}, {Rieke}, {Rieke}, {Robertson}, {Shivaei}, {Stark}, {Sun}, {Tacchella}, {Tang}, {Williams}, {Willmer}, {Baker}, {Baum}, {Bhatawdekar}, {Bowler}, {Boyett}, {Chen}, {Circosta}, {Helton}, {Ji}, {Kumari}, {Lyu}, {Nelson}, {Parlanti}, {Perna}, {Sandles}, {Scholtz}, {Suess}, {Topping}, {{\"U}bler}, {Wallace}, \& {Whitler}}]{2023A&A...677A..88B}
{Bunker}, A.~J., {Saxena}, A., {Cameron}, A.~J., {et~al.} 2023, \aap, 677, A88

\bibitem[{{Cameron} {et~al.}(2023){Cameron}, {Saxena}, {Bunker}, {D'Eugenio}, {Carniani}, {Maiolino}, {Curtis-Lake}, {Ferruit}, {Jakobsen}, {Arribas}, {Bonaventura}, {Charlot}, {Chevallard}, {Curti}, {Looser}, {Maseda}, {Rawle}, {Rodr{\'\i}guez Del Pino}, {Smit}, {{\"U}bler}, {Willott}, {Witstok}, {Egami}, {Eisenstein}, {Johnson}, {Hainline}, {Rieke}, {Robertson}, {Stark}, {Tacchella}, {Williams}, {Willmer}, {Bhatawdekar}, {Bowler}, {Boyett}, {Circosta}, {Helton}, {Jones}, {Kumari}, {Ji}, {Nelson}, {Parlanti}, {Sandles}, {Scholtz}, \& {Sun}}]{2023A&A...677A.115C}
{Cameron}, A.~J., {Saxena}, A., {Bunker}, A.~J., {et~al.} 2023, \aap, 677, A115

\bibitem[{{Centuri{\'o}n} {et~al.}(2003){Centuri{\'o}n}, {Molaro}, {Vladilo}, {P{\'e}roux}, {Levshakov}, \& {D'Odorico}}]{2003A&A...403...55C}
{Centuri{\'o}n}, M., {Molaro}, P., {Vladilo}, G., {et~al.} 2003, \aap, 403, 55

\bibitem[{{Cescutti} \& {Chiappini}(2010)}]{2010A&A...515A.102C}
{Cescutti}, G. \& {Chiappini}, C. 2010, \aap, 515, A102

\bibitem[{{Chabrier}(2003)}]{2003PASP..115..763C}
{Chabrier}, G. 2003, \pasp, 115, 763

\bibitem[{{Charbonnel} {et~al.}(2023){Charbonnel}, {Schaerer}, {Prantzos}, {Ram{\'\i}rez-Galeano}, {Fragos}, {Kuruvanthodi}, {Marques-Chaves}, \& {Gieles}}]{2023A&A...673L...7C}
{Charbonnel}, C., {Schaerer}, D., {Prantzos}, N., {et~al.} 2023, \aap, 673, L7

\bibitem[{{Chiappini} {et~al.}(2008){Chiappini}, {Ekstr{\"o}m}, {Meynet}, {Hirschi}, {Maeder}, \& {Charbonnel}}]{2008A&A...479L...9C}
{Chiappini}, C., {Ekstr{\"o}m}, S., {Meynet}, G., {et~al.} 2008, \aap, 479, L9

\bibitem[{{Chiappini} {et~al.}(2006){Chiappini}, {Hirschi}, {Meynet}, {Ekstr{\"o}m}, {Maeder}, \& {Matteucci}}]{2006A&A...449L..27C}
{Chiappini}, C., {Hirschi}, R., {Meynet}, G., {et~al.} 2006, \aap, 449, L27

\bibitem[{{Chiappini} {et~al.}(2005){Chiappini}, {Matteucci}, \& {Ballero}}]{2005A&A...437..429C}
{Chiappini}, C., {Matteucci}, F., \& {Ballero}, S.~K. 2005, \aap, 437, 429

\bibitem[{{Chieffi} \& {Limongi}(2013)}]{2013ApJ...764...21C}
{Chieffi}, A. \& {Limongi}, M. 2013, \apj, 764, 21

\bibitem[{{Christensen} {et~al.}(2012){Christensen}, {Laursen}, {Richard}, {Hjorth}, {Milvang-Jensen}, {Dessauges-Zavadsky}, {Limousin}, {Grillo}, \& {Ebeling}}]{2012MNRAS.427.1973C}
{Christensen}, L., {Laursen}, P., {Richard}, J., {et~al.} 2012, \mnras, 427, 1973

\bibitem[{{Curti} {et~al.}(2024){Curti}, {Witstok}, {Jakobsen}, {Kobayashi}, {Curtis-Lake}, {Hainline}, {Ji}, {D'Eugenio}, {Chevallard}, {Maiolino}, {Scholtz}, {Carniani}, {Arribas}, {Baker}, {Bhatawdekar}, {Boyett}, {Bunker}, {Cameron}, {Cargile}, {Charlot}, {Eisenstein}, {Ji}, {Johnson}, {Kumari}, {Maseda}, {Robertson}, {Silcock}, {Tacchella}, {Ubler}, {Venturi}, {Williams}, {Willmer}, \& {Willott}}]{2024arXiv240702575C}
{Curti}, M., {Witstok}, J., {Jakobsen}, P., {et~al.} 2024, Accepted for publication on A\&A, arXiv:2407.02575

\bibitem[{{D'Antona} {et~al.}(2023){D'Antona}, {Vesperini}, {Calura}, {Ventura}, {D'Ercole}, {Caloi}, {Marino}, {Milone}, {Dell'Agli}, \& {Tailo}}]{2023A&A...680L..19D}
{D'Antona}, F., {Vesperini}, E., {Calura}, F., {et~al.} 2023, \aap, 680, L19

\bibitem[{{D'Ercole} \& {Brighenti}(1999)}]{1999MNRAS.309..941D}
{D'Ercole}, A. \& {Brighenti}, F. 1999, \mnras, 309, 941

\bibitem[{{Ekstr{\"o}m} {et~al.}(2012){Ekstr{\"o}m}, {Georgy}, {Eggenberger}, {Meynet}, {Mowlavi}, {Wyttenbach}, {Granada}, {Decressin}, {Hirschi}, {Frischknecht}, {Charbonnel}, \& {Maeder}}]{2012A&A...537A.146E}
{Ekstr{\"o}m}, S., {Georgy}, C., {Eggenberger}, P., {et~al.} 2012, \aap, 537, A146

\bibitem[{{Frischknecht} {et~al.}(2016){Frischknecht}, {Hirschi}, {Pignatari}, {Maeder}, {Meynet}, {Chiappini}, {Thielemann}, {Rauscher}, {Georgy}, \& {Ekstr{\"o}m}}]{2016MNRAS.456.1803F}
{Frischknecht}, U., {Hirschi}, R., {Pignatari}, M., {et~al.} 2016, \mnras, 456, 1803

\bibitem[{{Fujita} {et~al.}(2004){Fujita}, {Mac Low}, {Ferrara}, \& {Meiksin}}]{2004ApJ...613..159F}
{Fujita}, A., {Mac Low}, M.-M., {Ferrara}, A., \& {Meiksin}, A. 2004, \apj, 613, 159

\bibitem[{{Grisoni} {et~al.}(2019){Grisoni}, {Matteucci}, {Romano}, \& {Fu}}]{2019MNRAS.489.3539G}
{Grisoni}, V., {Matteucci}, F., {Romano}, D., \& {Fu}, X. 2019, \mnras, 489, 3539

\bibitem[{{Hirschi}(2007)}]{2007A&A...461..571H}
{Hirschi}, R. 2007, \aap, 461, 571

\bibitem[{{Isobe} {et~al.}(2023){Isobe}, {Ouchi}, {Tominaga}, {Watanabe}, {Nakajima}, {Umeda}, {Yajima}, {Harikane}, {Fukushima}, {Xu}, {Ono}, \& {Zhang}}]{2023ApJ...959..100I}
{Isobe}, Y., {Ouchi}, M., {Tominaga}, N., {et~al.} 2023, \apj, 959, 100

\bibitem[{{Ji} {et~al.}(2024){Ji}, {{\"U}bler}, {Maiolino}, {D'Eugenio}, {Arribas}, {Bunker}, {Charlot}, {Perna}, {Rodr{\'\i}guez Del Pino}, {B{\"o}ker}, {Cresci}, {Curti}, {Kumari}, \& {Lamperti}}]{2024MNRAS.535..881J}
{Ji}, X., {{\"U}bler}, H., {Maiolino}, R., {et~al.} 2024, \mnras, 535, 881

\bibitem[{{Karakas}(2010)}]{2010MNRAS.403.1413K}
{Karakas}, A.~I. 2010, \mnras, 403, 1413

\bibitem[{{Kobayashi}(2022)}]{2022IAUS..366...63K}
{Kobayashi}, C. 2022, in IAU Symposium, Vol. 366, The Origin of Outflows in Evolved Stars, ed. L.~{Decin}, A.~{Zijlstra}, \& C.~{Gielen}, 63--82

\bibitem[{{Kobayashi} \& {Ferrara}(2024)}]{2024ApJ...962L...6K}
{Kobayashi}, C. \& {Ferrara}, A. 2024, \apjl, 962, L6

\bibitem[{{Kroupa}(2001)}]{2001MNRAS.322..231K}
{Kroupa}, P. 2001, \mnras, 322, 231

\bibitem[{{Kunth} {et~al.}(1995){Kunth}, {Matteucci}, \& {Marconi}}]{1995A&A...297..634K}
{Kunth}, D., {Matteucci}, F., \& {Marconi}, G. 1995, \aap, 297, 634

\bibitem[{{Limongi} \& {Chieffi}(2018)}]{2018ApJS..237...13L}
{Limongi}, M. \& {Chieffi}, A. 2018, \apjs, 237, 13

\bibitem[{{Mac Low} \& {Ferrara}(1999)}]{1999ApJ...513..142M}
{Mac Low}, M.-M. \& {Ferrara}, A. 1999, \apj, 513, 142

\bibitem[{{Maiolino} {et~al.}(2024){Maiolino}, {Scholtz}, {Witstok}, {Carniani}, {D'Eugenio}, {de Graaff}, {{\"U}bler}, {Tacchella}, {Curtis-Lake}, {Arribas}, {Bunker}, {Charlot}, {Chevallard}, {Curti}, {Looser}, {Maseda}, {Rawle}, {Rodr{\'\i}guez del Pino}, {Willott}, {Egami}, {Eisenstein}, {Hainline}, {Robertson}, {Williams}, {Willmer}, {Baker}, {Boyett}, {DeCoursey}, {Fabian}, {Helton}, {Ji}, {Jones}, {Kumari}, {Laporte}, {Nelson}, {Perna}, {Sandles}, {Shivaei}, \& {Sun}}]{2024Natur.627...59M}
{Maiolino}, R., {Scholtz}, J., {Witstok}, J., {et~al.} 2024, \nat, 627, 59

\bibitem[{{Marconi} {et~al.}(1994){Marconi}, {Matteucci}, \& {Tosi}}]{1994MNRAS.270...35M}
{Marconi}, G., {Matteucci}, F., \& {Tosi}, M. 1994, \mnras, 270, 35

\bibitem[{{Marques-Chaves} {et~al.}(2024){Marques-Chaves}, {Schaerer}, {Kuruvanthodi}, {Korber}, {Prantzos}, {Charbonnel}, {Weibel}, {Izotov}, {Messa}, {Brammer}, {Dessauges-Zavadsky}, \& {Oesch}}]{2024A&A...681A..30M}
{Marques-Chaves}, R., {Schaerer}, D., {Kuruvanthodi}, A., {et~al.} 2024, \aap, 681, A30

\bibitem[{{Martin} {et~al.}(2002){Martin}, {Kobulnicky}, \& {Heckman}}]{2002ApJ...574..663M}
{Martin}, C.~L., {Kobulnicky}, H.~A., \& {Heckman}, T.~M. 2002, \apj, 574, 663

\bibitem[{{Matteucci}(1986)}]{1986MNRAS.221..911M}
{Matteucci}, F. 1986, \mnras, 221, 911

\bibitem[{{Matteucci}(2021)}]{2021A&ARv..29....5M}
{Matteucci}, F. 2021, \aapr, 29, 5

\bibitem[{{Matteucci} {et~al.}(1997){Matteucci}, {Molaro}, \& {Vladilo}}]{1997A&A...321...45M}
{Matteucci}, F., {Molaro}, P., \& {Vladilo}, G. 1997, \aap, 321, 45

\bibitem[{{Meynet} \& {Maeder}(2002{\natexlab{a}})}]{2002A&A...390..561M}
{Meynet}, G. \& {Maeder}, A. 2002{\natexlab{a}}, \aap, 390, 561

\bibitem[{{Meynet} \& {Maeder}(2002{\natexlab{b}})}]{2002A&A...381L..25M}
{Meynet}, G. \& {Maeder}, A. 2002{\natexlab{b}}, \aap, 381, L25

\bibitem[{{Milakovi{\'c}} {et~al.}(2024){Milakovi{\'c}}, {Webb}, {Molaro}, {Lee}, {Jethwa}, {Cupani}, {Murphy}, {Welsh}, {D'Odorico}, {Cristiani}, {G{\'e}nova Santos}, {Martins}, {Nunes}, {Schmidt}, {Pepe}, {Zapatero Osorio}, {Alibert}, {Gonz{\'a}lez Hern{\'a}ndez}, {Di Marcantonio}, {Palle}, {Rebolo}, {Santos}, {Sousa}, \& {Su{\'a}rez Mascare{\~n}o}}]{2024MNRAS.534...12M}
{Milakovi{\'c}}, D., {Webb}, J.~K., {Molaro}, P., {et~al.} 2024, \mnras, 534, 12

\bibitem[{{Molaro}(2003)}]{2003ASPC..304..221M}
{Molaro}, P. 2003, in Astronomical Society of the Pacific Conference Series, Vol. 304, CNO in the Universe, ed. C.~{Charbonnel}, D.~{Schaerer}, \& G.~{Meynet}, 221

\bibitem[{{Molaro}(2006)}]{2006cams.book..256M}
{Molaro}, P. 2006, in Chemical Abundances and Mixing in Stars in the Milky Way and its Satellites, ed. S.~{Randich} \& L.~{Pasquini}, 256

\bibitem[{{Molaro} {et~al.}(2023){Molaro}, {Aguado}, {Caffau}, {Allende Prieto}, {Bonifacio}, {Gonz{\'a}lez Hern{\'a}ndez}, {Rebolo}, {Zapatero Osorio}, {Cristiani}, {Pepe}, {Santos}, {Alibert}, {Cupani}, {Di Marcantonio}, {D'Odorico}, {Lovis}, {Martins}, {Milakovi{\'c}}, {Murphy}, {Nunes}, {Schmidt}, {Sousa}, {Sozzetti}, \& {Su{\'a}rez Mascare{\~n}o}}]{2023A&A...679A..72M}
{Molaro}, P., {Aguado}, D.~S., {Caffau}, E., {et~al.} 2023, \aap, 679, A72

\bibitem[{{Molaro} {et~al.}(2004){Molaro}, {Centuri{\'o}n}, {D'Odorico}, \& {P{\'e}roux}}]{2004oee..sympE..39M}
{Molaro}, P., {Centuri{\'o}n}, M., {D'Odorico}, V., \& {P{\'e}roux}, C. 2004, in Origin and Evolution of the Elements, ed. A.~{McWilliam} \& M.~{Rauch}, 39

\bibitem[{{Molaro} {et~al.}(1996){Molaro}, {D'Odorico}, {Fontana}, {Savaglio}, \& {Vladilo}}]{1996A&A...308....1M}
{Molaro}, P., {D'Odorico}, S., {Fontana}, A., {Savaglio}, S., \& {Vladilo}, G. 1996, \aap, 308, 1

\bibitem[{{Molero} {et~al.}(2023){Molero}, {Magrini}, {Matteucci}, {Romano}, {Palla}, {Cescutti}, {Viscasillas V{\'a}zquez}, \& {Spitoni}}]{2023MNRAS.523.2974M}
{Molero}, M., {Magrini}, L., {Matteucci}, F., {et~al.} 2023, \mnras, 523, 2974

\bibitem[{{M{\"u}ller} {et~al.}(2016){M{\"u}ller}, {Heger}, {Liptai}, \& {Cameron}}]{2016MNRAS.460..742M}
{M{\"u}ller}, B., {Heger}, A., {Liptai}, D., \& {Cameron}, J.~B. 2016, \mnras, 460, 742

\bibitem[{{Nagele} \& {Umeda}(2023)}]{2023ApJ...949L..16N}
{Nagele}, C. \& {Umeda}, H. 2023, \apjl, 949, L16

\bibitem[{{Nandal} {et~al.}(2024){Nandal}, {Sibony}, \& {Tsiatsiou}}]{2024A&A...688A.142N}
{Nandal}, D., {Sibony}, Y., \& {Tsiatsiou}, S. 2024, \aap, 688, A142

\bibitem[{{Noterdaeme} {et~al.}(2017){Noterdaeme}, {Krogager}, {Balashev}, {Ge}, {Gupta}, {Kr{\"u}hler}, {Ledoux}, {Murphy}, {P{\^a}ris}, {Petitjean}, {Rahmani}, {Srianand}, \& {Ubachs}}]{2017A&A...597A..82N}
{Noterdaeme}, P., {Krogager}, J.~K., {Balashev}, S., {et~al.} 2017, \aap, 597, A82

\bibitem[{{Ott} {et~al.}(2005){Ott}, {Walter}, \& {Brinks}}]{2005MNRAS.358.1453O}
{Ott}, J., {Walter}, F., \& {Brinks}, E. 2005, \mnras, 358, 1453

\bibitem[{{Palicio} {et~al.}(2024){Palicio}, {Matteucci}, {Della Valle}, \& {Spitoni}}]{2024A&A...689A.203P}
{Palicio}, P.~A., {Matteucci}, F., {Della Valle}, M., \& {Spitoni}, E. 2024, \aap, 689, A203

\bibitem[{{Pascale} {et~al.}(2023){Pascale}, {Dai}, {McKee}, \& {Tsang}}]{2023ApJ...957...77P}
{Pascale}, M., {Dai}, L., {McKee}, C.~F., \& {Tsang}, B. T.~H. 2023, \apj, 957, 77

\bibitem[{{Pettini} {et~al.}(1995){Pettini}, {Lipman}, \& {Hunstead}}]{1995ApJ...451..100P}
{Pettini}, M., {Lipman}, K., \& {Hunstead}, R.~W. 1995, \apj, 451, 100

\bibitem[{{Pilyugin}(1993)}]{1993A&A...277...42P}
{Pilyugin}, L.~S. 1993, \aap, 277, 42

\bibitem[{{Prantzos} {et~al.}(2018){Prantzos}, {Abia}, {Limongi}, {Chieffi}, \& {Cristallo}}]{2018MNRAS.476.3432P}
{Prantzos}, N., {Abia}, C., {Limongi}, M., {Chieffi}, A., \& {Cristallo}, S. 2018, \mnras, 476, 3432

\bibitem[{{Recchi} {et~al.}(2001){Recchi}, {Matteucci}, \& {D'Ercole}}]{2001MNRAS.322..800R}
{Recchi}, S., {Matteucci}, F., \& {D'Ercole}, A. 2001, \mnras, 322, 800

\bibitem[{{Recchi} {et~al.}(2008){Recchi}, {Spitoni}, {Matteucci}, \& {Lanfranchi}}]{2008A&A...489..555R}
{Recchi}, S., {Spitoni}, E., {Matteucci}, F., \& {Lanfranchi}, G.~A. 2008, \aap, 489, 555

\bibitem[{{Renzini} \& {Voli}(1981)}]{1981A&A....94..175R}
{Renzini}, A. \& {Voli}, M. 1981, \aap, 94, 175

\bibitem[{{Rivera-Thorsen} {et~al.}(2024){Rivera-Thorsen}, {Chisholm}, {Welch}, {Rigby}, {Hutchison}, {Florian}, {Sharon}, {Choe}, {Dahle}, {Bayliss}, {Khullar}, {Gladders}, {Hayes}, {Adamo}, {Owens}, \& {Kim}}]{2024A&A...690A.269R}
{Rivera-Thorsen}, T.~E., {Chisholm}, J., {Welch}, B., {et~al.} 2024, \aap, 690, A269

\bibitem[{{Rizzuti} {et~al.}(2019){Rizzuti}, {Cescutti}, {Matteucci}, {Chieffi}, {Hirschi}, \& {Limongi}}]{2019MNRAS.489.5244R}
{Rizzuti}, F., {Cescutti}, G., {Matteucci}, F., {et~al.} 2019, \mnras, 489, 5244

\bibitem[{{Rizzuti} {et~al.}(2021){Rizzuti}, {Cescutti}, {Matteucci}, {Chieffi}, {Hirschi}, {Limongi}, \& {Saro}}]{2021MNRAS.502.2495R}
{Rizzuti}, F., {Cescutti}, G., {Matteucci}, F., {et~al.} 2021, \mnras, 502, 2495

\bibitem[{{Roberti} {et~al.}(2024){Roberti}, {Limongi}, \& {Chieffi}}]{2024ApJS..270...28R}
{Roberti}, L., {Limongi}, M., \& {Chieffi}, A. 2024, \apjs, 270, 28

\bibitem[{{Romano} {et~al.}(2020){Romano}, {Franchini}, {Grisoni}, {Spitoni}, {Matteucci}, \& {Morossi}}]{2020A&A...639A..37R}
{Romano}, D., {Franchini}, M., {Grisoni}, V., {et~al.} 2020, \aap, 639, A37

\bibitem[{{Romano} {et~al.}(2019){Romano}, {Matteucci}, {Zhang}, {Ivison}, \& {Ventura}}]{2019MNRAS.490.2838R}
{Romano}, D., {Matteucci}, F., {Zhang}, Z.-Y., {Ivison}, R.~J., \& {Ventura}, P. 2019, \mnras, 490, 2838

\bibitem[{{Rossi} {et~al.}(2024){Rossi}, {Romano}, {Mucciarelli}, {Ceccarelli}, {Massari}, \& {Zamorani}}]{2024A&A...691A.284R}
{Rossi}, M., {Romano}, D., {Mucciarelli}, A., {et~al.} 2024, \aap, 691, A284

\bibitem[{{Salpeter}(1955)}]{1955ApJ...121..161S}
{Salpeter}, E.~E. 1955, \apj, 121, 161

\bibitem[{{Schaerer} {et~al.}(2024){Schaerer}, {Marques-Chaves}, {Xiao}, \& {Korber}}]{2024A&A...687L..11S}
{Schaerer}, D., {Marques-Chaves}, R., {Xiao}, M., \& {Korber}, D. 2024, \aap, 687, L11

\bibitem[{{Senchyna} {et~al.}(2024){Senchyna}, {Plat}, {Stark}, {Rudie}, {Berg}, {Charlot}, {James}, \& {Mingozzi}}]{2024ApJ...966...92S}
{Senchyna}, P., {Plat}, A., {Stark}, D.~P., {et~al.} 2024, \apj, 966, 92

\bibitem[{{Smartt} {et~al.}(2009){Smartt}, {Eldridge}, {Crockett}, \& {Maund}}]{2009MNRAS.395.1409S}
{Smartt}, S.~J., {Eldridge}, J.~J., {Crockett}, R.~M., \& {Maund}, J.~R. 2009, \mnras, 395, 1409

\bibitem[{{Stiavelli} {et~al.}(2024){Stiavelli}, {Morishita}, {Chiaberge}, {Leethochawalit}, {Norman}, {Ricotti}, {Roberts-Borsani}, {Treu}, {Vanzella}, {Wyse}, {Zhang}, \& {Boyett}}]{2024arXiv241206517S}
{Stiavelli}, M., {Morishita}, T., {Chiaberge}, M., {et~al.} 2024, Submitted to ApJ, arXiv:2412.06517

\bibitem[{{Sukhbold} {et~al.}(2016){Sukhbold}, {Ertl}, {Woosley}, {Brown}, \& {Janka}}]{2016ApJ...821...38S}
{Sukhbold}, T., {Ertl}, T., {Woosley}, S.~E., {Brown}, J.~M., \& {Janka}, H.~T. 2016, \apj, 821, 38

\bibitem[{{Tapia} {et~al.}(2024){Tapia}, {Bekki}, \& {Groves}}]{2024MNRAS.534.2086T}
{Tapia}, T., {Bekki}, K., \& {Groves}, B. 2024, \mnras, 534, 2086

\bibitem[{{Vangioni} {et~al.}(2018){Vangioni}, {Dvorkin}, {Olive}, {Dubois}, {Molaro}, {Petitjean}, {Silk}, \& {Kimm}}]{2018MNRAS.477...56V}
{Vangioni}, E., {Dvorkin}, I., {Olive}, K.~A., {et~al.} 2018, \mnras, 477, 56

\bibitem[{{Ventura} {et~al.}(2013){Ventura}, {Di Criscienzo}, {Carini}, \& {D'Antona}}]{2013MNRAS.431.3642V}
{Ventura}, P., {Di Criscienzo}, M., {Carini}, R., \& {D'Antona}, F. 2013, \mnras, 431, 3642

\bibitem[{{Villar-Mart{\'\i}n} {et~al.}(2004){Villar-Mart{\'\i}n}, {Cervi{\~n}o}, \& {Gonz{\'a}lez Delgado}}]{2004MNRAS.355.1132V}
{Villar-Mart{\'\i}n}, M., {Cervi{\~n}o}, M., \& {Gonz{\'a}lez Delgado}, R.~M. 2004, \mnras, 355, 1132

\bibitem[{{Vincenzo} {et~al.}(2016){Vincenzo}, {Belfiore}, {Maiolino}, {Matteucci}, \& {Ventura}}]{2016MNRAS.458.3466V}
{Vincenzo}, F., {Belfiore}, F., {Maiolino}, R., {Matteucci}, F., \& {Ventura}, P. 2016, \mnras, 458, 3466

\bibitem[{{Watanabe} {et~al.}(2024){Watanabe}, {Ouchi}, {Nakajima}, {Isobe}, {Tominaga}, {Suzuki}, {Ishigaki}, {Nomoto}, {Takahashi}, {Harikane}, {Hatano}, {Kusakabe}, {Moriya}, {Nishigaki}, {Ono}, {Onodera}, \& {Sugahara}}]{2024ApJ...962...50W}
{Watanabe}, K., {Ouchi}, M., {Nakajima}, K., {et~al.} 2024, \apj, 962, 50

\bibitem[{{Welsh} {et~al.}(2020){Welsh}, {Cooke}, {Fumagalli}, \& {Pettini}}]{2020MNRAS.494.1411W}
{Welsh}, L., {Cooke}, R., {Fumagalli}, M., \& {Pettini}, M. 2020, \mnras, 494, 1411

\bibitem[{{Woosley} \& {Weaver}(1995)}]{1995ApJS..101..181W}
{Woosley}, S.~E. \& {Weaver}, T.~A. 1995, \apjs, 101, 181

\bibitem[{{Zafar} {et~al.}(2014){Zafar}, {Centuri{\'o}n}, {P{\'e}roux}, {Molaro}, {D'Odorico}, {Vladilo}, \& {Popping}}]{2014MNRAS.444..744Z}
{Zafar}, T., {Centuri{\'o}n}, M., {P{\'e}roux}, C., {et~al.} 2014, \mnras, 444, 744

\end{thebibliography}

\end{document}